# Algorithmic Verification of Single-Pass List Processing Programs

Rajeev Alur[1] and Pavol Černý[2]

[1] University of Pennsylvania
[2] IST Austria

**Abstract.** We introduce *streaming data string transducers* that map input data strings to output data strings in a single left-to-right pass in linear time. Data strings are (unbounded) sequences of data values, tagged with symbols from a finite set, over a potentially infinite data domain that supports only the operations of equality and ordering. The transducer uses a finite set of states, a finite set of variables ranging over the data domain, and a finite set of variables ranging over data strings. At every step, it can make decisions based on the next input symbol, updating its state, remembering the input data value in its data variables, and updating data-string variables by concatenating data-string variables and new symbols formed from data variables, while avoiding duplication. We establish that the problems of checking functional equivalence of two streaming transducers, and of checking whether a streaming transducer satisfies pre/post verification conditions specified by streaming acceptors over input/output data-strings, are in PSPACE.

We identify a class of imperative and a class of functional programs, manipulating lists of data items, which can be effectively translated to streaming data-string transducers. The imperative programs dynamically modify a singly-linked heap by changing next-pointers of heap-nodes and by adding new nodes. The main restriction specifies how the next-pointers can be used for traversal. We also identify an expressively equivalent fragment of functional programs that traverse a list using syntactically restricted recursive calls. Our results lead to algorithms for assertion checking and for checking functional equivalence of two programs, written possibly in different programming styles, for commonly used routines such as insert, delete, and reverse.

## 1 Introduction

We propose streaming transducers as an abstract and analyzable model for programs that access and modify sequences of data items in a single pass. The idea of using transducers for modeling such programs is a natural one. However, the class of regular transductions, which has appealing theoretical properties such as MSO characterization, is defined by two-way transducers. As an example, consider the *reverse* transduction that reverses the input string. It is not definable using classical one-way transducers (such as Mealy machines), we need a second (backward) pass during which the output is produced. On the other hand, this transduction can easily be computed by a single-pass program that traverses a list both in the settings of imperative programs manipulating heap-allocated lists and functional programs using tail recursion. Our streaming transducer model can capture such computations naturally. Furthermore, we show that verification problems such as checking functional equivalence, assertion checking, and checking correctness with respect to pre/post conditions, are decidable for this transducer model, even in the presence of values from an unbounded data domain.

In the proposed model, a (deterministic) *streaming data-string transducers* map input data strings to output data strings in a single left-to-right pass in linear time. A *data string* is a sequence of items of type `data × tag`, where `data` is a potentially infinite set of data values, and `tag` is a finite set of labels. The only operations allowed on the type `data` are tests for equality and ordering, and this restriction is essential for decidability. The transducer uses a finite set of states, a finite set of variables ranging over `data`, and a finite set of variables ranging over data strings. At every step, it can make decisions based on the current state, the tag of the next input symbol, and the ordering relationship of the data of the next input symbol with the data values stored in data variables. It can update the state, modify data variables using the input data value, and update the data-string variables using assignments whose right-hand-sides are concatenations of data-string variables and new symbols formed using data variables. A key restriction is that a data-string variable can be used at most once in a right-hand-side expression at each step. Multiple data-string variables are necessary for the transducer to compute different possible chunks of the output, and the restriction on how they can be used ensures that at every step there is merely a rearrangement of outputs computed so far without duplication.

We consider the following two decision problems for streaming transducers: (1) *Equivalence:* given two streaming transducers, do they define the same (partial) functions? (2) *Pre-post condition checking:* given a streaming transducer, a pre-condition and a post-condition, both expressed by similarly defined streaming acceptors over data strings, is it the case that for every input satisfying the pre-condition, the corresponding transducer output satisfies the post-condition? We show both problems to be in PSPACE. We also show that extending the model along several possible directions leads quickly to undecidability of basic problems such as reachability.

We then identify a class of programs that precisely correspond to the streaming data-string transducers. The input to a program is a single list with elements of type `data × tag`, possibly with additional arguments of types `data`, `tag`, and `bool`, and the output is a single list, possibly with additional returned values of types `data`, `tag`, and `bool`. The key restriction, needed for decidability of verification problems, is that the program computes the output in a single pass processing the next item of the list at each step. A number of commonly used routines such as insertion, deletion, membership, reversal, sorting with respect to tags (but not data values), naturally satisfy this restriction.

For heap-manipulating imperative programs, the input list is stored in a heap of nodes each of which can store a tag, a data value, and a next-node pointer. During the computation of the program, the next-pointers induce an unranked forest structure over the nodes. The program accesses the heap using a finite number of pointer variables and uses a finite number of data variables. The program can add new nodes to the heap, change values stored at nodes referenced by pointer variables, and also modify next-pointers of such nodes. A key restriction on the traversal of the heap using next-pointers is that the only legal use of the next-field on the right-hand-side is in the assignment *curr* := *curr.next*, where *curr* is the unique pointer initially pointing to the head of the input list. We also show that for this class of programs, a variety of assertion checking problems (such as "is a program location reachable," and "does the heap stay acyclic") are solvable in PSPACE.

Finally, we present a class of list-processing functional programs which traverse an input list from left to right using recursive calls. The key restriction is that a call to the function $f$ with input list $l$ can recursively call $f$ with input `tail l`, and returns value obtained by composing its input arguments and the values returned by the recursive call without examining them. We show that this class precisely corresponds to the streaming data string transducers. Thus, the results of this paper show how to automatically compile two list-processing programs, one written in imperative style, and one written in functional style, into an intermediate low-level transducer model, and algorithmically check if the two are semantically equivalent.

## 2 Streaming transducers

### 2.1 Data strings and transductions

A *data domain* is a totally ordered, possibly infinite, set $D$ of data values. We will use $<$ to denote the strict total order over $D$. Throughout this paper, assume $D$ to be fixed. A *data symbol* over $\Sigma$, where $\Sigma$ is a finite set of symbols or *tags*, is a pair $(\sigma, d)$ with $\sigma \in \Sigma$ and $d \in D$. A *data string* $w$ over $\Sigma$ is a finite sequence $(\sigma_1, d_1), (\sigma_2, d_2), \ldots (\sigma_k, d_k)$ of data symbols over $\Sigma$. A *data language* over $\Sigma$ is a set $L$ of data strings over $\Sigma$. A *(deterministic) data transduction* from an input alphabet $\Sigma$ to an output alphabet $\Gamma$ is a partial function $F$ from data strings over $\Sigma$ to data strings over $\Gamma$. For a data transduction $F$ from an input alphabet $\Sigma$ to an output alphabet $\Gamma$, the *domain* of $F$ is the data language over $\Sigma$ consisting of data strings $w$ such that $F(w)$ is defined.

As an example, let $D$ be the set of strings over ASCII characters ordered by the standard lexicographic ordering. Let $\Sigma$ contain two tags *private* and *public*. A data symbol denotes an entry in an address-book consisting of a name tagged either as private or public. A data string represents an address-book. Here are a few examples of data languages: language $L_1$ consists of all data strings in which names appear in the alphabetic order (that is, data symbols are sorted in an increasing order according to $<$ over data values); and language $L_2$ consists of all data strings that do not contain duplicate entries. A few examples of useful data transductions are: transduction $F_1$ maps a data string to its reverse; transduction $F_2$ maps a data string $w$ to $w_1.w_2$, where $w_1$ is the subsequence of $w$ containing private entries, and $w_2$ is the subsequence of $w$ containing public entries; and transduction $F_3$ deletes an entry (*private*,$d$) from the input data string if the string also contains (*public*,$d$). To model operations such as insertion and deletion that take data values/symbols as inputs in addition to a data string, we can encode all inputs in a single data string. For example, the transduction $F_4$, given an input data string $(\sigma, d).w$ checks if names appear in the alphabetic order in the tail $w$, and if so, it returns $w$ with $(\sigma, d)$ inserted in the correct position to maintain the output string sorted ($F_4$ is undefined if the input string is empty, or if the tail of the input string is not sorted).

## 2.2 Transducer definition

We now describe our model of deterministic transducers. The transducer reads an input data string left-to-right in a single pass, and computes an output data string. The transducer uses a finite set of states, a finite set of data variables that range over data values, and a finite set of data string variables that range over data strings over the output alphabet. At each step, the transducer reads the next data symbol of the input string, and chooses a transition depending on the current state, the tag of the input symbol, and the ordering relationship of the data value of the input symbol with values of all its data variables. The transition updates the state, updates the data variables possibly using the input data value, and updates the data string variables in parallel using assignments whose right-hand-sides are concatenations of data string variables and new data symbols formed using data variables. When the transducer consumes the entire input string, the final output string is produced by similarly concatenating data string and data variables. A key restriction is that a data string variable can be used at most once in a right-hand-side expression in a transition, and thus, at every step, there is merely a rearrangement of output chunks computed so far, without duplication.

We now define the model formally. A *(deterministic) streaming data-string transducer* (SDST) $S$ from an input alphabet $\Sigma$ to an output alphabet $\Gamma$ consists of a finite set of states $Q$, an initial state $q_0 \in Q$, a finite set of data variables $V$, a data variable $curr \in V$ used to refer to the data value of the current input symbol, a finite set of data string variables $X$, a partial output function $O$ from $Q$ to $((\Gamma \times V) \cup X)^*$, a finite set $E$ of transitions of the form $(q, \sigma, \varphi, q', \alpha)$, where $q \in Q$ is a source state, $\sigma \in \Sigma$ is an input tag, $\varphi$ is a Boolean formula over atomic constraints of the form $v < curr$ and $curr < v$ with $v \in V$, $q' \in Q$ is a target state, and $\alpha$ is an assignment mapping data variables $V$ to $V$ and data string variables $X$ to $((\Gamma \times V) \cup X)^*$. It is required that (1) for each $q \in Q$ and $x \in X$, there is at most one occurrence of $x$ in $O(q)$, and (2) for each transition $(q, \sigma, \varphi, q', \alpha)$, for each $x \in X$, $x$ appears at most once in the set of strings $\{\alpha(y) \mid y \in X\}$, and (3) for each pair of transitions $(q, \sigma, \varphi, q', \alpha)$ and $(q, \sigma, \varphi', q'', \alpha')$ with the same source state and input tag, the tests $\varphi$ and $\varphi'$ are mutually exclusive (that is, $\varphi \wedge \varphi'$ is unsatisfiable).

A *valuation* $\beta$ for such a transducer $S$ is a *partial* function over data and data string variables such that for each data variable $v \in V$, either $\beta(v)$ is undefined or is a data value in $D$, and for each data string variable $x \in X$, either $\beta(x)$ is undefined or is a data string over the output alphabet $\Gamma$. Such a valuation naturally extends to a partial Boolean function to evaluate tests. Each test $\varphi$ is a Boolean formula over atomic constraints of the form $v < curr$ and $curr < v$ with $v \in V$. The value $\beta(\varphi)$ is defined if $\beta(v)$ is defined for all data variables $v$ occurring in $\varphi$, and if so, $\varphi$ is evaluated using the values $\beta$ assigns to these data variables. A valuation $\beta$ also extends to strings in $((\Gamma \times V) \cup X)^*$: given a string $u$ in $((\Gamma \times V) \cup X)^*$, the valuation $\beta(u)$ is defined when $\beta$ is defined for all the data and data string variables occurring in $u$, and if so, $\beta(u)$ is the data string over the output alphabet $\Gamma$ obtained by replacing each data string variable $x$ in $u$ with the data string $\beta(x)$ and each data variable $v$ in $u$ with the data value $\beta(v)$.

Given an SDST $S$, a *configuration* of $S$ is a pair $(q, \beta)$, where $q$ is a state in $Q$, and $\beta$ is a valuation for $S$. The initial configuration is $(q_0, \beta_0)$ where $q_0$ is the initial state of $S$, $\beta_0(v)$ is undefined for each data variable $v$, and $\beta_0(x)$ is the empty string for each data string variable $x$. The one-step transition relation over the set of configurations is defined as follows. Consider a configuration $(q, \beta)$ and an input data symbol $(\sigma, d)$. The transducer first updates the valuation $\beta$ to $\beta'$ by setting *curr* to the input data value $d$. Next, let $(q, \sigma, \varphi, q', \alpha)$ be a transition such that $\beta'$ satisfies the test $\varphi$. If there is such a transition, then the updated state is $q'$, and the updated value of each data and data string variable $x$ is obtained by evaluating the right-hand-side $\alpha(x)$ according to the valuation $\beta'$. That is, if there exists a transition $(q, \sigma, \varphi, q', \alpha)$ such that $\beta'(\varphi) = 1$, for $\beta' = \beta[curr \mapsto d]$, then $(q, \beta) \xrightarrow{\sigma, d} (q', \beta' \cdot \alpha)$. Determinism ensures that each configuration has at most one successor for a given input data symbol. The transition relation extends to a multi-step relation over input data strings in the natural way. Given an input data string $w$ over $\Sigma$, $(q_0, \beta_0) \xrightarrow{w} (q, \beta)$ means that the configuration of the transducer after reading the input data string $w$ is $(q, \beta)$ (if no such configuration exists that means no transition is enabled at some step). The semantics of $S$ is then defined to be the transduction $[\![S]\!]$ defined as: for an input data string $w$ over $\Sigma$, if $(q_0, \beta_0) \xrightarrow{w} (q, \beta)$ and $O(q)$ is defined then $[\![S]\!](w)$ is defined to be $\beta(O(q))$, otherwise $[\![S]\!](w)$ is undefined.

We call a data transduction $F$ from an input alphabet $\Sigma$ to an output alphabet $\Gamma$ to be *streaming-regular* if there exists an SDST $S$ such that $[\![S]\!] = F$.

### 2.3 Examples

To illustrate our definition of transducers, let us consider the transductions mentioned in Section 2.1. The transduction $F_1$ to reverse the input data string can be implemented by a streaming data-string transducer $S_1$ with a single state, a single data variable *curr*, and a single data string variable $x$. The input tag $\sigma$ is processed by the self-loop transition with update $x := (\sigma, curr).x$ (by default, a variable that is not explicitly updated, remains unchanged; we omit such assignments for readability). The output function outputs $x$. No tests on input data values are needed. Notice that classical definitions of transducers allow adding output symbols only at the end of the output computed so far. Adding a symbol to the front of the string variable $x$ at each step is crucial to implement reverse in a single left-to-right pass.

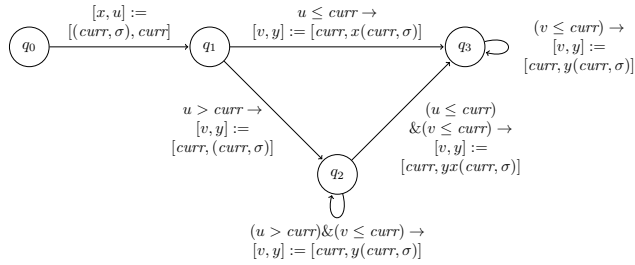

**Fig. 1.** Transduction $F_4$

Now let us consider the transduction $F_2$ that maps a data string $w$ to $w_1.w_2$, where $w_1$ and $w_2$ are the subsequences of $w$ containing private and public entries, respectively. This can be implemented by an SDST $S_2$ that maintains two data string variables $x_1$ and $x_2$, and a single data variable $curr$. At each step, if the tag of current input symbol is *private*, the symbol is added to $x_1$ (the precise assignment is $[x_1, x_2] := [x_1.(private, curr), x_2]$), otherwise the symbol is added to $x_2$ in a symmetric manner. The output function outputs the concatenation $x_1.x_2$. Note that it is not possible to implement this transduction by an SDST using just one variable.

The transduction $F_3$ deletes a private entry from the input data string if the string also contains a public entry with a matching data value. When reading an input symbol with data value $d$, the streaming algorithm needs to figure out if a private entry with the same data value has been encountered before. An SDST with $k$ variables can effectively use only $k$ data values for tests at any step, and since the number of possible data values in an input string is unbounded, $F_3$ is not streaming-regular.

Consider the transduction $F_4$ that inserts the head symbol of the input string in its tail, provided that the tail is sorted. The transducer $S_4$ uses three data variables: $u$ to remember the head data value, $v$ to remember the previous data value, and $curr$ to refer to the current data value. It uses a data string variable $x$ to remember the first data symbol, and $y$ to compute the output. The transducer is shown in Figure 1. The transducer is in state $q_0$ initially, in state $q_1$ after reading one symbol, in state $q_2$ after reading 2 or more symbols provided the tail so far is sorted and all its data values are smaller than the head data value (stored in $u$), and in state $q_3$ after reading 2 or more symbols provided the tail so far is sorted and the head symbol has already been inserted in the output. In states $q_2$ and $q_3$, the variable $v$ stores the previous data value, and the test $v \leq curr$ checks for sortedness of the input ($v$ gets updated to $curr$ at each step). If this test does not hold, no transition is enabled, and the output will be undefined. The transition to $q_3$ inserts the data symbol stored in $x$ in the output $y$. The output function is undefined in state $q_0$, and is $x$ in state $q_1$, $y.x$ in state $q_2$, and $y$ in state $q_3$.

### 2.4 Streaming acceptors

A *streaming data-string acceptor* (SDSA) is a streaming data-string transducer $S$ with an empty set of data string variables. Such an acceptor $S$ has a finite set of data variables that can remember the data values from the input string, and can make decisions based on their relative ordering. The output alphabet plays no role in the behavior of an acceptor. Given an input data string $w$, either the output $[\![S]\!](w)$ is defined or undefined, and the domain of the transducer is the data language associated with the acceptor $S$. This is the same as saying that the output function $O$ marks states of $S$ as accepting or rejecting based on whether the output function $O$ is defined or undefined at a state. We call a data language $L$ over an alphabet $\Sigma$ to be *streaming-regular* if $L$ is accepted by a streaming data-string acceptor.

The data language $L_1$ consisting of sorted data strings can be defined by such an acceptor using one data variable that remembers the previous data value, along with the data variable *curr* needed to refer to the data value of the currently read symbol. Thus $L_1$ is a streaming-regular data language. The data language $L_2$ consisting of data strings without duplicate entries is not streaming-regular by an argument similar to the one for the transduction $F_3$.

Among different types of automata over data strings that have been studied, *data automata* [3] have emerged as a good candidate for the notion of regularity for languages of data strings. However, data automata are too expressive for our purpose as they have an undecidable emptiness problem in the presence of ordering on data values [3].

### 2.5 Properties

In this section, we note some properties of streaming transducers aimed at understanding their expressiveness. First, observe that a streaming data-string transducer $S$ cannot output "new" data values. That is, for every input data string $w$, any data value appearing in the output data string $[\![S]\!](w)$ must appear in some symbol in $w$. Second, streaming transducers are *bounded* in the sense that the length of the output string is within at most a constant factor of the length of the input string.

**Proposition 1.** *If $F$ is a streaming-regular transduction from $\Sigma$ to $\Gamma$, then for all input data strings $w$ over $\Sigma$, $|F(w)| = O(|w|)$.*

The boundedness depends on the fact that the parallel assignment at each step is copyless: each variable can appear in a right-hand-side expression at most once. Not only this is crucially needed for decidability of the equivalence problem, it also allows an efficient implementation: if the data strings corresponding to variables are stored in linked lists, then the assignment can be executed by only changing a constant number of pointers (proportional to the description of the transducer, but independent of the lengths of the data strings they store, and thus, independent of the length of the input string).

**Proposition 2.** *If $F$ is a streaming-regular data-string transduction, then given an input string $w$, the output $F(w)$ can be computed in time $O(|w|)$.*

This also means that the sorting transduction that maps an input data string to its sorted version is not streaming-regular due to well-known lower bounds for sorting. Obviously streaming transducers cannot capture all linear-time streaming algorithms. As a specific example, let us revisit the transduction $F_2$ that maps a data string $w$ to $w_1.w_2$, where $w_1$ and $w_2$ are the projections of $w$ containing private and public entries, respectively. Consider the variation $F_2'$ that maps $w$ to a *merge* of the two projections $w_1$ and $w_2$ taking elements from the two lists in an alternate manner. This can be easily implemented in linear-time if we maintain two read-heads over the input string, one corresponding to private entries and one corresponding to public entries. Note that the emptiness problem of finite automata with multiple read-heads is undecidable, and the traversal allowed for streaming transducers is restricted by design to ensure decidability of key analysis problems. In particular, $F_2'$ is not streaming-regular.

Let us now consider some closure properties for the class of streaming-regular transducers. Given two data transductions $F_1$ and $F_2$, and a test $L$ as a data language, suppose we want to compute $F_1(w)$ when $w \in L$ and $F_2(w)$ otherwise. If all of $F_1$, $F_2$, and $L$ are specified using SDSTs then we can construct an SDST for the desired transduction by a suitably modified product construction.

**Proposition 3.** *If $F_1$ and $F_2$ are streaming-regular data transductions from $\Sigma$ to $\Gamma$, and $L$ is a streaming-regular data language over $\Sigma$, then the following data transduction $F$ is streaming-regular: for an input data string $w$ over $\Sigma$, if $w \in L$ then $F(w) = F_1(w)$ else $F(w) = F_2(w)$.*

It turns out that streaming-regular data transductions are not closed under functional composition. That is, given two SDSTs $S_1$ and $S_2$, we cannot always construct an SDST $S$ such that $S(w) = S_2(S_1(w))$.

**Proposition 4.** *There exist streaming-regular data transductions $F_1$ and $F_2$ from $\Sigma$ to $\Sigma$ such that the following data transduction $F$ is not streaming-regular: for an input data string $w$ over $\Sigma$, $F(w) = F_2(F_1(w))$.*

*Proof.* We choose $\Sigma$ to be a singleton set, and thus, it plays no role. Consider the transduction $F_1$ that maps a data string $d_1 d_2 \cdots d_k$ to its reverse $d_k d_{k-1} \cdots d_1$. $F_1$ is streaming-regular. Consider the transduction $F_2$ that maps a data string $d_1 d_2 \cdots d_k$ to $(d_1)^k$. That is, $F_2$ just repeats the first data value for each input symbol read. It is easy to implement $F_2$ by an SDST. Now consider the composition $F = F_1 \cdot F_2$. The transduction $F$ maps an input data string $d_1 d_2 \cdots d_k$ to $(d_k)^k$. We can prove that $F$ is not streaming-regular.

Note that the above proof crucially uses the fact that the data domain $D$ is unbounded, and we can always find a "fresh" data value that has not appeared in the input string before. If we make $D$ finite, then the streaming-regular transducers are closed under composition.

## 3 Imperative Programs Updating Linked Lists

We consider a class of imperative programs that manipulate heap-allocated singly-linked list data structures. Each node of the heap stores a tag, a data value, and a pointer to another node. For clarity, in this section, we will assume that the output alphabet is the same as the input alphabet, so we need to consider tags of only one type. The input data string is stored in such a heap using one node for each position (null pointer indicates the end of the list). A list-processing program is invoked with the reference to the head-node of the list as input. The program traverses the list using next-pointers, and computes using variables that range over tags, over data values, over booleans, and over pointers into the heap. It can create new nodes and add them to the heap, and can also manipulate the shape of the heap by updating the next-pointers of the nodes referenced by its pointer variables. The output data string is returned using a pointer-variable that points to the head of the list storing that output. During the computation of the program, next-pointers of two heap-nodes may point to the same node, and thus, the heap in general has a structure of an unordered forest. Since the output is computed by possibly reusing the nodes that store the input, we need careful syntactic restrictions to allow a single-pass traversal of the input list, while disallowing repeated or nested traversals. We require that a typical traversal assignment $x := y.next$ for pointer variables $x$ and $y$ is disallowed. The only legal use of the next-field on the right-hand-side is in the assignment $curr := curr.next$, where $curr$ is the unique input pointer. Assignments of the form $x.next := y$ to update the heap structure are allowed, provided $x$ and $curr$ are not referencing the same heap-node. An attempt to execute $x.next := y$ in a state where $x$ and $curr$ reference the same heap-node, causes a runtime error (alternatively we can require each such assignment to be syntactically guarded by the boolean condition $x \neq curr$).

A program can have additional input and output parameters, and each such input/output parameter can be a data value, a boolean value, or a tag. Before we describe the syntax and the semantics in detail, let us first consider a couple of examples. The following function reverses a list in-place, and corresponds to the data transduction $F_1$:

```
function Reverse
  input ref curr;   output ref result := curr;
  local ref prev := curr;
  if curr ≠ nil then {
    curr := curr.next;
    while curr ≠ nil {
      result := curr;   curr := curr.next;
```

      result.next := prev;   prev := result;
  } }.

Suppose given an input data string $w$ and an input data symbol $d$, we want to delete each symbol in $w$ whose data value equals $d$, and return the resulting data string along with a boolean flag that indicates whether or not some symbol was actually deleted. The following function implements this:

```
function Delete
  input ref curr;   input data v;
  output ref result;   output bool b := 0;
  local ref prev;
  while (curr ≠ nil) & (curr.data = v) {
    curr := curr.next;   b := 1;
  }
  result := curr;   prev :=curr;
  if curr ≠ nil then {
    curr := curr.next;   prev.next := nil;
    while curr ≠ nil {
      if curr.data = v then {
        curr := curr.next;   b := 1;
      } else {
        prev.next := curr;   prev := curr;
        curr := curr.next;   prev.next := nil;
    }; }; }.
```

### 3.1 Syntax

*Types:* Variables are typed. The possible types are: `bool` for Boolean-valued variables, `tag` for variables ranging over the alphabet $\Sigma$, `data` for variables ranging over the data domain $D$ along with an "undefined" value denoted $\perp$, and `ref` for reference variables that index into the data heap along with the null reference `nil`.

*Variable declarations:* A program variable is declared along with its type (`bool`, `tag`, `data`, or `ref`) and an annotation which can be either `local`, `input`, or `output`. The `input` annotation means that the variable is an input to the function. A function has exactly one input variable of type `ref`, and can have multiple input variables of other types. We will use *curr* to name this unique input reference variable. The `output` annotation means that the variable is an output of the function, and `local` annotation means that the variable is neither an input nor an output. There is exactly one output variable of type `ref` which is used to return a single data string. The declaration of each output and local variable has an associated value. The initial value of a variable of type `bool` or `tag` can be either a constant or an input variable of matching type. The initial value of a data variable can be either $\bot$ or an input data variable. The initial value of a pointer variable can be either *curr* or `nil`.

*Data expressions and assignments:* A data expression is of the form (1) a variable of type `data`, or (2) *r.data*, where *r* is a variable of type `ref`, denoting the data value stored in the heap-node indexed by *r*. A data assignment statement assigns a data expression to a data variable.

*Tag expressions and assignments:* A tag expression is of the form (1) a variable of type `tag`, (2) a constant $\sigma$ from the alphabet $\Sigma$, or (3) *r.tag*, where *r* is a variable of type `ref`, denoting the tag value stored in the heap-node indexed by *r*. A tag assignment statement assigns a tag expression to a tag variable.

*Reference expressions and assignments:* A reference expression *re* is either a variable of type `ref` or the constant `nil`. A reference assignment statement is either (1) $r := re$, where *r* is a local or a output variable of type `ref` and *re* is a reference expression, (2) $r.next := re$, where *r* is of type `ref` and *re* is a reference expression, (3) $r := \text{new}(te, de, re)$, where *r* is a local or a output variable of type `ref` and *te* is a tag expression, *de* is a data expression, *re* is a reference expression, or (4) *curr := curr.next*. The first assignment allows reassignment of reference variables, except for the input variable *curr*. The second assignment updates the heap by changing the next-pointer of the heap-node indexed by *r*, provided *r* and *curr* do not point to the same heap-node. The third assignment creates a new heap-node with tag value given by *te*, data value given by *de*, and next-pointer given by *re*. The last assignment allows traversal, and is syntactically restricted to ensure that only the unique input reference variable is used to traverse the input list.

*Boolean expressions and assignments:* An atomic boolean expression is either a boolean constant (0/1), or tests equality between two tag expressions, or tests equality or ordering between two data expressions, or tests equality between two reference expressions. A boolean expression is formed from atomic boolean expressions using standard logical connectives for negation, conjunction, and disjunction. A boolean assignment statement assigns a boolean expression to a boolean variable.

*Statements:* An assignment statement is either a data assignment statement, a tag assignment statement, a reference assignment statement, or a boolean assignment statement. A statement *s* is either (1) an assignment statement, (2) a conditional statement of the form `if` *be* `then` *s* or `if` *be* `then` $s_1$ `else` $s_2$, where *be* is a boolean expression, (3) a while statement of the form `while` *be* `{` *s*`}`, where *be* is a boolean expression, or (4) a finite sequence of statements.

*Program:* A *single-pass list processing program P* consists of a sequence of variable declarations followed by a statement.

### 3.2 Semantics

Recall that a program has a single input variable of type `ref` and a single output variable of type `ref`. The semantics of a program is defined as a partial function from an input data string together with values for input data/tag/boolean variables to an output data string together with values for output data/tag/boolean variables. For example, the semantics of `Delete` is a partial function from $(\Sigma \times D)^* \times D$ to $(\Sigma \times D)^* \times \{0, 1\}$.

*Configurations:* Given a program $P$, its configuration is completely described by (1) the values of its data, tag, boolean, and reference variables, (2) the program counter indicating the next statement to be executed, and (3) the data-heap. Let *Loc* be the set of locations in $P$ (this can be the set of vertices in the control-flow graph of the program). A data-heap $h$ consists of a finite set $N$ of heap-nodes, a data function $f_d : N \mapsto D$ that gives the data element stored at each node, a tag function $f_t : N \mapsto \Sigma$ that gives the tag element stored at each node, and a next-pointer function $f_n : N \mapsto N_\perp$ that gives the next-pointer of each node, where $N_\perp$ is the set $N$ together with the constant `nil`. A program-configuration $c$ of $P$ then consists of a location $\ell \in Loc$, a data heap $h = (N, f_d, f_t, f_n)$, and a partial function $\beta$ over all the program variables that maps each data variable to $D$, each reference variable to $N$, each boolean variable to $\{0, 1\}$, and each tag variable to $\Sigma$.

*Initialization:* Given an input data string $(\sigma_1, d_1) \cdots (\sigma_k, d_k)$, the initial heap $h_0$ consists of the set $N = \{n_1, \ldots n_k\}$ of nodes, one per each data symbol of the input string. The data function is given by $f_d(n_i) = d_i$, the tag function is given by $f_t(n_i) = \sigma_i$, and the next-pointer function is given by $f_n(n_i) = n_{i+1}$ for $i < k$ and $f_n(n_k) = \mathtt{nil}$. The initial location $\ell_0$ is the unique entry location of the control-flow graph. For the initial valuation $\beta_0$, $\beta_0(curr) = n_1$. For all other input variables $x$, $\beta_0(x)$ is set to the corresponding input value. For all local and output variables $x$, $\beta_0(x)$ is defined according to the initialization in the declaration for $x$. The initial configuration $c_0$ of the program is $(\ell_0, h_0, \beta_0)$.

*Transition relation over configurations:* The operational semantics of programs is defined by a transition relation over the configurations. First, given a configuration $c = (\ell, (N, f_d, f_t, f_n), \beta)$, there is a natural way to evaluate a data expression $de$ to obtain a data value $c(de) \in D$, a tag expression $te$ to obtain a tag value $c(te) \in \Sigma$, a reference expression $re$ to obtain a value $c(re) \in N_\perp$, and a boolean expression $be$ to obtain a boolean value $c(be)$. Every program configuration $c = (\ell, (N, f_d, f_t, f_n), \beta)$ can have at most one successor configuration, determined by the statement $s$ at location $\ell$. The details are standard, and we illustrate them using a few cases.

Suppose the statement is a conditional statement $\ell : \mathtt{if}\ b\ \mathtt{then}\ \ell_1 : s_1\ \mathtt{else}\ \ell_2 : s_2$. Then, if $c(b) = 1$ then the successor configuration of $c$ is $(\ell_1, h, \beta)$, and if $c(b) = 0$ then the successor configuration of $c$ is $(\ell_2, h, \beta)$.

Suppose the statement $s$ is a reference assignment statement $\ell : \ r := \mathtt{new}(te, de, re)$. The effect of executing the statement $s$ updates the control location from $\ell$ to the unique successor location $\ell'$ of the statement $s$. For the updated data heap $h'$, the set of nodes is $N \cup \{n\}$, where $n \notin N$ is a "new" heap-node, the data function is $f_d[n \mapsto c(de)]$, the tag function is $f_t[n \mapsto c(re)]$, and the next-pointer function is $f_n[n \mapsto c(re)]$. The updated valuation $\beta'$ is $\beta[r \mapsto n]$.

Suppose the statement $s$ is a reference assignment statement $\ell : \ r.next := re$. If $c(r) = c(curr)$ then this is an error and the configuration $c$ has no successor. Otherwise, the successor configuration is $c'$ such that the location $\ell'$ is the unique successor location of $\ell$ in the control-flow graph, the valuation $\beta$ stays unchanged, and the updated heap is $(N, f_d, f_t, f_n[c(r) \mapsto c(re)])$ (that is, the next-pointer of the node $c(r)$ in the heap changes to $c(re)$ which may be $\mathtt{nil}$ or a heap-node).

*Termination and output:* An execution of the program is obtained by starting in the initial configuration $c_0$ and continuing with the successor configuration as long as possible. If this execution is infinite, then the program is non-terminating, and the output is undefined. Suppose the execution is finite and ends in the configuration $c_f = (\ell_f, h_f, \beta_f)$. If the location $\ell_f$ is not the unique exit location of the control-flow graph, then again the output is undefined. Otherwise, the returned value of each output data/tag/boolean variable is given by the final valuation $\beta_f$ of program variables. For the unique output reference variable $r$, let $(\sigma_1, d_1)(\sigma_2, d_2)\cdots$ be the unique sequence of tag/data values stored in the heap $h_f$ starting at the node $\beta_f(r)$ following the next-pointers until the $\mathtt{nil}$ value is encountered. If this sequence is infinite, this indicates that the program created a cycle in the heap during its computation, and the output is again undefined. If this sequence is finite, it is the returned output data string.

### 3.3 Streaming transducers with $\epsilon$-transitions

We extend the model of streaming data-string transducers by allowing the transducer to update its state, data variables, and data string variables using an $\epsilon$-transition that does not consume an input symbol. We will first show that it is possible to eliminate such $\epsilon$-transitions, and then we will translate list-processing programs to transducers with $\epsilon$-transitions.

The definition of a (deterministic) streaming data-string transducer $S$ with $\epsilon$-transitions extends the definition of SDSTs as follows: in a transition $(q, \sigma, \varphi, q', \alpha)$, $\sigma$ can now also be $\epsilon$, provided there is no transition of the form $(q, \sigma', \varphi', q'', \alpha')$ with $\sigma' \in \Sigma$. The restriction is needed for ensuring determinism: in a state $q$, either all outgoing transitions are $\epsilon$-transitions, or all outgoing transitions have non-$\epsilon$ tags (and thus consume the next input symbol). Note that the original determinism requirement still applies: if there are multiple transitions with same source state and same input tag (which now may be $\epsilon$), their tests must be mutually exclusive.

As in case of SDSTs, a configuration consists of a state $q$ and a valuation $\beta$ for the data and data string variables. The definition of the transition relation $(q, \beta) \xrightarrow{\sigma, d} (q', \beta')$, for $\sigma \in \Sigma$ and $d \in D$, is unchanged. The $\epsilon$-transitions are defined by: if there exists a transition $(q, \epsilon, \varphi, q', \alpha)$ such that $\beta(\varphi) = 1$ then $(q, \beta) \xrightarrow{\epsilon} (q', \beta \cdot \alpha)$. A run over the input data string $w$ is obtained by starting in the initial configuration $(q_0, \beta_0)$, and applying either an $\epsilon$-transition or a transition corresponding to the next input data symbol until all the input data symbols are consumed and no more $\epsilon$-transitions are possible. This ensures determinism: for a given input string $w$, there is at most one configuration $(q, \beta)$ such that (1) $(q_0, \beta_0) \xrightarrow{w} (q, \beta)$ and (2) $(q, \beta)$ has no $\epsilon$-successor. The semantics $[\![S]\!](w)$ is defined to be $\beta(O(w))$ in such a case, provided $O(q)$ is defined, and is undefined otherwise. Note that it is possible that such a transducer keeps on executing $\epsilon$-transitions without terminating, and in such a case, the corresponding output is undefined.

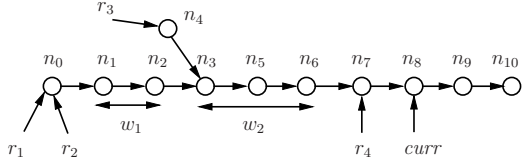

**Fig. 2.** Storing heap in data strings

It turns out this extension does not add to the expressiveness:

**Proposition 5.** *Given a streaming data-string transducer $S$ with $\epsilon$-transitions, one can effectively construct a streaming data-string transducer (without $\epsilon$-transitions) $S'$ such that $[\![S]\!] = [\![S']\!]$ with the same number of states, same number of data variables, and the same number of data string variables.*

### 3.4 From single-pass programs to streaming transducers

In this section, we describe how to translate single-pass list processing programs to streaming data-string transducers. The first step is to view the semantics of a single-pass list processing program as a partial function from data strings to data strings. To associate such a data string transduction $[\![P]\!]$ with a program $P$, we encode input parameters in the same manner as described in Sec. 2.1. If $P$ has $k_i$ input boolean/data/tag variables and $k_o$ output boolean/data/tag variables, then we prefix the input data string with $k_i$ symbols each encoding one input argument, and prefix the output data string with $k_o$ symbols each encoding one output value.

The main challenge in the construction is to store the information in the data heap used by the program $P$ using a bounded number of data and data string variables in the corresponding transducer $S$. Figure 2 shows a possible configuration of the data heap that the program accesses using the reference variables $curr$ and $r_1, r_2, r_3, r_4$. The first observation is that the heap-nodes that are not accessible from any of the reference variables are not relevant to the execution of the program, and can be ignored. Second, nodes such as $n_9$ and $n_{10}$ that are accessible from *curr.next* can contain only input symbols that the program has not processed so far. These nodes have not influenced the execution of the program so far, and information in these nodes does not need to be stored. When the program executes the statement *curr* := *curr.next*, the node $n_9$ becomes relevant. This step is analogous to the transducer $S$ processing the next input symbol.

Compressing the heap using a bounded number of strings is achieved using an encoding similar to [14]. A node is called a referenced node if a reference variable points to it. In the example, $n_0$, $n_4$, $n_7$ and $n_8$ are referenced nodes. Information in such nodes needs to be stored explicitly by $S$. For each reference variable $r$ of $P$, $S$ maintains a data variable $d_r$ and a tag variable $t_r$ storing the information in the node that $r$ points to. A node such as $n_3$ is called an interruption node as two nodes point to it (and both these nodes are accessible from the program's reference variables). If $P$ has $k$ reference variables, then there can be at most $2k - 1$ interruption nodes. The stretches $n_1, n_2$ and $n_3, n_5, n_6$ are uninterrupted heap segments. In each such segment (1) the first node is either an interruption node, or is the next-successor of a referenced node, (2) the next-pointer of each node in the sequence points to the next node in the sequence, (3) no node other than the first is an interruption or a referenced node, and (4) the next-pointer of the last node is either nil or points to an interruption or a referenced node. If $P$ has $k$ reference variables, then there can be at most $2k - 1$ uninterrupted heap segments. The sequence of data symbols stored in an uninterrupted heap segment is stored in a data string variable by $S$. In our example, the data string $w_1$ stores the data symbols in $n_1, n_2$ and the data string $w_2$ stores the data symbols in $n_3, n_4, n_5$. The finite-state control of $S$ remembers the *shape* of the heap: $r_1$ and $r_2$ point to the same node, the next-pointer of the $r_1$-referenced node points to the data string stored in $w_1$, the next-pointers of the node referenced by $r_3$ and of the last node of $w_1$ point to $w_2$, $w_2$ is followed by the $r_4$-referenced node, which is followed by the *curr*-referenced node. Such a shape can be captured by a function $f_n : Y \mapsto Y_\perp$, where $Y$ contains all the reference variables of $P$ and all the data string variables of $S$ that store the data strings in uninterrupted heap segments. If $P$ executes the assignment $r_3.next := curr$, then $n_3$ is no longer an interruption node, and in this case, the two uninterrupted heap segments collapse into one. This is achieved by $S$ by the assignment $[w_1, w_2] := [w_1.w_2, \varepsilon]$ to the data string variables, updating the tag/data variable corresponding to $r_3$, and changing the shape by updating $f_n(w_1)$ to $r_4$ and $f_n(r_3)$ to *curr*.

**Proposition 6.** *Given a single-pass list-processing program $P$ one can effectively construct a streaming data-string transducer $S$ with $\epsilon$-transitions such that $[\![P]\!] = [\![S]\!]$. If $P$ has $m$ locations, $k_r$ reference variables, $k_b$ boolean variables, $k_t$ tag variables, and $k_d$ data variables, then $S$ has $k_d + k_r$ data variables, $2k_r$ data string variables, and $O(m \cdot 2^{k_b} \cdot k_r^{k_r} \cdot |\Sigma|^{k_t + k_r})$ states.*

*Proof.* The transducer $S$ has a data variable for each data variable for $P$, and also for each reference variable of $P$ (to store the data values in referenced nodes in the heap). It has $2k_r$ data string variables to store uninterrupted heap segments. The state of $S$ stores (1) the location of control of $P$, (2) the boolean value of each boolean variable of $P$, (3) the tag value of each tag variable of $P$, (4) a partition of the reference variables of $P$ (two reference variables are in the same partition if they point to the same heap-node), (5) for each equivalence class in the partition, either a tag value stored at the node referenced, or `nil`, and (6) for each data string variable and each equivalence class of the partition, a next value that gives either a data string variable or an equivalence class of the partition. The last component stores the shape of the heap. The bound on the possible number of states follows by a simple counting argument.

If $P$ has data, boolean, or tag input variables, the transducer $S$ first scans the initial prefix of data symbols setting up the initial values using $\epsilon$-transitions. Then the transducer processes the first input symbol, assigning the tag-component corresponding to *curr* in its state to tag of the input symbol, and assigning the data variable corresponding to *curr* to data of the input symbol. After this phase, the control location is set to the unique entry location, all output strings are empty (there are no uninterrupted heap segments in the initial heap), and the partition has only two classes (some reference variables are `nil` and some are in the class that contains *curr*).

We now describe transitions of $S$ corresponding to statements of $P$. The only statement that causes a non-$\epsilon$ transition is the statement $curr := curr.next$. The transition that corresponds to this statement in $S$: (1) changes the stored control location of $P$, (2) changes the partition of reference variables into equivalence classes: *curr* is split from its current equivalent class, (3) a new tag value is stored for the new equivalence class of *curr*, (4) for each reference variable $p$ from the previous equivalence class of *curr* the new next value of $p$ (i.e. $f_n(p)$) will be *curr*, (5) if for any data string variable $x$ (that stores an uninterrupted heap segment) of $S$ we have $(f_n(x) = curr)$ then we append the current data symbol (from before the transition is executed) to $x$. All other statements are captured by an $\epsilon$-transition, as they do not correspond to the move of the head of the automaton. Boolean, tag, and data assignments can be simulated directly. We have already described using an example how a statement $r.next = curr$ can affect the shape of the heap, and how this is captured by assignments that $S$ can perform in its transitions.

In Section 2.5, we have mentioned that the assignments that a streaming transducer performs on its data string variables can be executed by only changing a constant number of pointers. A list-processing program equivalent to a transducer stores data strings in list segments on the heap, and keeps pointers to the first and last nodes of the segment. To perform an assignment of the form $x := xy$, the program performs commands $xl.next := yf; xl := yl$, where $xf$ ($xl$) is a reference to the first (last) node representing $x$ (and similarly for $y$). We obtain the following proposition:

**Proposition 7.** *Given a streaming data string transducer $S$, one can effectively construct a single-pass list-processing program $P$ such that $[\![S]\!] = [\![P]\!]$.*

## 4 Functional programs on lists

We consider a simply typed functional language with types `bool`, `tag`, `data`, list, and function types, with `letrec` recursion and pair and list constructors. As in the previous section, we assume that $\Sigma = \Gamma$ for ease of presentation. The terms are defined by:

```
t:=    true | false | isTrue t | isFalse t
    | S | isS (for all S ∈ tag)
    | x | fun x:T.t | t t | if t then t else t
    | x = x | x < x | {t,t} | t.1 | t.2
    | nil | cons t t | append t t
    | isnil t | head t | tail t
    | let x=t in t | letrec x:T=t in t
```

The operators `=` and `<` apply to terms of type `data`. The lists are of type (`list` (`tag` × `data`)), or `list` for short. The list and pair terms are standard.

We defined a class of functional programs that intuitively captures single-pass functions that process a list by recursing through it from left to right. The recursion restriction we define is a minor generalization of tail recursive functions, where we allow the caller to perform operations on, but not test, the values that the callee returns. This allows capturing common routines such as insert, delete, and reverse (both tail-recursive and non-tail recursive). We allow wrapper functions in order to enable the standard programming style for tail recursive functions.

A function is a *single-pass list processing function* iff it is defined by the following term:

```
fun a:list × T1.
  letrec f:((list × T1 × T2) → (list × T)) = t
  in (f a initValues)
```

This term encodes a wrapper function that can pass some additional arguments to a recursive function `f`. The following conditions are required to hold:

- *one list*: Let us denote the first argument to `f` (by definition of type `list`) by `l`. The list `l` is the only list accessed by `isnil`, `head` and `tail`. The contents of the other list variables are thus not examined. However, it is possible to use `cons` or `append` with these variables.
- *type restrictions*: The types `T1` and `T` are products of types `bool`, `tag`, `data`, with possibly more than one component of each type. These are input and output arguments of the list processing function. The type `T2` is a product of types `bool`, `tag`, `data`, and `list`, with possibly more than one component of each type. Intuitively, these are buffers where the recursive function can store results.
- *recursion restriction*: The first argument to any recursive call is the term (`tail l`), where `l` is the first argument to `f`. Every recursive call to `f` is enclosed in an expression `e` defined by `let r =(f (tail l) a) in t`. Furthermore, `e` is the last expression the caller evaluates, i.e. `t` is the caller's result. Using the type restriction above, we have that $\mathtt{r} \equiv \mathtt{l}_1, \mathtt{r}_1, \ldots, \mathtt{r}_n$, and $\mathtt{a} \equiv \mathtt{a}_1 \ldots \mathtt{a}_n$, and $\mathtt{t} \equiv \mathtt{l}_2, \mathtt{t}_1, \ldots \mathtt{t}_n$, where $\mathtt{l}_1$ is a list and $\mathtt{l}_2$ is a list expression, each of $\mathtt{r}_1, \ldots \mathtt{r}_n$ and $\mathtt{t}_1, \ldots \mathtt{t}_n$ is of type `bool`, `tag` or `data`, and each of $\mathtt{a}_1, \ldots \mathtt{a}_n$ is of type `bool`, `tag`, `data`, or `list`. We have the following restrictions on these subexpressions: (i) a list variable of `f` can appear in at most one expression $\mathtt{a}_i$ or $\mathtt{t}_j$ (this is similar to the restriction that requires the assignments of SDSTs to be "copyless"), (ii) if $\mathtt{t}_i$ is of type `bool`, then the only variable from `r` it can use is $\mathtt{r}_i$, (iii) if $\mathtt{t}_i$ is of type `tag` or `data`, then we have $\mathtt{t}_i = \mathtt{r}_i$, (iv) the only variable from `r` that $\mathtt{l}_2$ can use is $\mathtt{l}_1$.
- *term t*: The term `t` is of the form `fun x: (list × T1 × T2). t'`, where $\mathtt{t}'$ does not contain the function definition term `fun` or the recursive definition term `letrec`.
- *initValues*: The expression `initValues` is of type `T2`. If `T2` contains `list` types, the corresponding values in `initValues` are `nil`.

Note that all the above conditions can be checked syntactically.

### 4.1 Examples

The data transduction $F_1$ that reverses a list can be encoded as a single-pass list processing function as follows:

```
fun l:list.
  letrec reverse: (list × list → list)
      = fun l:list. fun result:list.
    if (isnil l) then result
    else (reverse (tail l) (cons (head l) result))
  in (reverse l nil)
```

A function that given a list $l$ and a data value $d$ deletes all occurrences of $d$ in $l$ is encoded as follows:

```
fun l:list. d:data.
  letrec FuncDelete: (list × data → list)
      = fun l:list. fun d:data.
    if (isnil l) then nil
    else
      if (head l).data = d
        then (FuncDelete (tail l) d)
        else (cons (head l)
                   (FuncDelete (tail l) d))
  in (FuncDelete l d)
```

One of the recursive calls in `FuncDelete` is enclosed in an expression that adds a cell to the front of the list: `(cons (head l) (FuncDelete (tail l) d))`. This recursive call is not tail recursive, as the caller function applies an operation to the result returned by the callee. However, the recursive call satisfies the recursion restriction from our definition of single-pass list processing functions (and it satisfies the other restrictions as well).

The constructions in this paper lead to an algorithm to check if the imperative function Delete of Section 3 and the above function FuncDelete are semantically equivalent (i.e., specify the same transduction). Such a check can be used for full functional verification of one using the other as the specification.

### 4.2 From list processing functions to streaming transducers

*Semantics* The simply-typed functional language we defined contains standard terms. The values of the language, as well as the typing rules, and the evaluation relation $t_1 \to t_2$ for all the terms are omitted here, as they can be found in a standard textbook [16]. The operational semantics is given by a transition system, whose states are subterms of f, whose transitions are given by the evaluation relation $t_1 \to t_2$, and whose initial state is the term f.

As for imperative programs, the semantics of a single-pass list processing function can be viewed as a data string transduction, that is, a partial function from data strings to data strings. To associate a transduction $[\![f]\!]$ with a single-pass list processing function $f : (\text{list} \times T_1) \to T$, we encode input parameters in the same way as in Section 3. Given a data string $w$, its first $k_i$ symbols can represent $k_i$ input boolean/data/tag variables, and the rest represents the input list (converting a data string to a list term of type list is straightforward). Given a data string, the function $decode\_param(w, k_i)$ returns the parameter values, and the function $decode\_list(w, k_i)$ returns the tail of the input string represented as a list term. Similar encoding to data strings can be used for output values and output lists. Given a tuple $\{l_1, r_1, \ldots, r_n\}$ of type $\text{list} \times T$, $enc\_res(\{l_1, r_1, \ldots, r_n\})$ returns the corresponding data string. Given a single-pass list processing function f, we have that $[\![f]\!](w) = w'$ iff $(f\ decode\_list(w)\ decode\_param(w)) \to^* \{l_1, r_1, \ldots, r_n\}$ and $enc\_res(l_1, r_1, \ldots, r_n) = w$.

**Proposition 8.** *Given a single-pass list processing function* $f : (\text{list} \times T_1) \to T$, *one can effectively construct a streaming data string transducer $S$, such that* $[\![f]\!] = [\![S]\!]$. *Let* g *be the recursive function used by* f. *If* g *has $k_b$ boolean variables, $k_t$ tag variables, $k_d$ data variables, and $k_l$ list variables, then $S$ has $O(2^{k_b} \cdot |\Sigma|^{k_t})$ states, $k_d$ data variables, and $k_l + 1$ data string variables.*

We note that the construction used in the proof of Proposition 8 is more direct than the one used in the proof of Proposition 6. The list variables (apart from the list that is traversed) and data variables are modeled directly by data string variables and data variables of the transducer, and the control state of the transducer encodes the value of boolean and tag variables.

Given an SDST $S$, we can construct an equivalent list-processing program f. We first describe the arguments of the recursive function g the function f uses: its boolean arguments encode state of $S$, its data arguments correspond to data variables of $S$, and its list arguments correspond to the data string variables of $S$. A transition $(q, \sigma, \varphi, q', \alpha)$ is translated by making the function $g$ test whether the current boolean arguments correspond to $q$, whether the current tag is $\sigma$, and whether $\varphi$ holds for the current data arguments. If so, the function makes a recursive call with parameters encoding $q'$ and the assignments from $\alpha$. The function we obtain in this way is tail recursive. The next proposition follows:

**Proposition 9.** *Given a single-pass data string transducer $S$, one can effectively construct a single-pass list processing function* $\mathtt{f}$, *such that* $[\![S]\!] = [\![\mathtt{f}]\!]$.

## 5 Decision Problems

In this section, we prove that the equivalence problem and the pre/post condition checking problem are decidable for streaming data string transducers. We also show that, for a number of extensions of the streaming transducer model, already the basic analysis problem of reachability is undecidable.

### 5.1 Sound and complete abstraction for order and equality

In proofs of decidability of equivalence and pre/post condition checking of SDSTs that operate on an infinite data domain $D$, we will construct finite state systems that do not store values of the data variables of the SDSTs, but only keep track of order and equality predicates. In order to prove that such an abstraction is both sound and complete for analysis problems, we will need the lemma presented in this subsection.

Let $V$ be a set of variables that range over $D$. We fix $V$ and an infinite $D$ for this subsection. We will consider pairs of the form $(V^d, \rho)$, where the set $V^d \subseteq V$ represents the set of variables with a defined value, and where $\rho$ is an *ec-order* on $V^d$ (short for order on equivalence classes). An ec-order $\rho = (\equiv_\rho, <_\rho)$ is a pair where the first component is an equivalence relation on $V^d$, and the second component is a strict total order on equivalence classes of $\equiv_\rho$. For data variables $v_1, v_2$, we write $v_1 <_\rho v_2$, if $v_1$ belongs to an equivalence class $c_1$, $v_2$ belongs to an equivalence class $c_2$, and $c_1 <_\rho c_2$. For example, if $V = \{v_1, v_2, v_3\}$, all variables have a defined value, then a possible ec-order on $V^d$ can be represented as $v_1 \equiv_\rho v_3 <_\rho v_2$. Let $\beta$ be a valuation of data variables as in the definition of SDSTs. A pair $(V^d, \rho)$ represents a set of valuations. We write $\beta \models (V^d, \rho)$ iff $\beta$ is defined precisely for the variables in $V^d$, and for all $v_1, v_2 \in V^d$ we have that $\beta(v_1) < \beta(v_2)$ iff $v_1 <_\rho v_2$, and $\beta(v_1) = \beta(v_2)$ iff $v_1 \equiv_\rho v_2$. Let $\varphi$ be a Boolean combination of constraints of the form $v_1 < v_2$ and $v_1 = v_2$ for variables $v_1, v_2 \in V$. Let $\alpha$ be a map from $V$ to $V$ modeling assignments, as in the definition of transitions of SDSTs. We write $\beta_1 \xrightarrow{(\varphi, \alpha)} \beta_2$, if $\beta_1$ satisfies $\varphi$ and $\beta_2 = \beta_1 \cdot \alpha$, similarly to the definition of SDSTs. For pairs $(V^d, \rho)$, we define a transition relation $(V^d, \rho) \xrightarrow{(\varphi, \alpha)} (V'^d, \rho')$ iff (a) $V'^d$ contains the variables which were assigned to by $\alpha$ from variables in $V^d$, (b) $\rho$ implies $\varphi$, and (c) $\rho'$ is the ec-order obtained from $\rho$ by executing $\alpha$. Let $u^a$ be a sequence of pairs $(V_1^d, \rho_1)(V_2^d, \rho_2) \ldots (V_n^d, \rho_n)$. Let $u$ be a sequence of valuations $\beta_1 \beta_2 \ldots \beta_n$. Let $upd$ be a sequence of pairs $(\varphi_1, \alpha_1)(\varphi_2, \alpha_2) \ldots (\varphi_{n-1}, \alpha_{n-1})$. The sequence $u^a$ *conforms* to the sequence $upd$ if for all $i$ if $1 \leq i < n$, then $(V_i^d, \rho_i) \xrightarrow{(\varphi_i, \alpha_i)}$

$(V_{i+1}^d, \rho_{i+1})$. Similarly, the sequence $u$ *conforms* to the sequence $upd$ if for all $i$ if $1 \leq i < n$, then $\beta_i \xrightarrow{(\varphi_i, \alpha_i)} \beta_{i+1}$.

The proof of the following lemma crucially uses the fact that the infinite totally-ordered data domain $D$ contains chains, that is, sequences of elements in an increasing order, of unbounded length. The proof is omitted here in the interest of space. A similar proof is a part of the proof of Theorem 1 of [2].

**Lemma 1.** *Let $upd$ be a sequence of pairs $(\varphi_1, \alpha_1)(\varphi_2, \alpha_2) \ldots (\varphi_{n-1}, \alpha_{n-1})$. Let $u^a$ be a sequence of pairs $(V_1^d, \rho_1)(V_1^d, \rho_1) \ldots (V_n^d, \rho_n)$, such that $u^a$ conforms to $upd$. Then there exists a sequence of valuations $u = \beta_1 \beta_2 \ldots \beta_n$ such that $u$ conforms to $upd$ and for all $i$, if $1 \leq i \leq n$, then $u_i \models u_i^a$.*

### 5.2 Equivalence checking

Given two streaming data-string transducers $S_1$ and $S_2$ from $\Sigma$ to $\Gamma$, the *streaming transducer equivalence problem* is to determine whether $[\![S_1]\!] = [\![S_2]\!]$.

In order to show that the problem can be solved in PSPACE we reduce the problem to a reachability problem in 1-counter machines. A 1-counter machine $M$ is a tuple $(Q_M, \delta_M, q_0^M, F_M)$, where $Q_M$ is a set of states, $q_0^M$ is the initial state, and $F_M \subseteq Q_M$ is a set of final states. The transition relation $\delta_M$ is a relation in $Q_M \times Q_M \times \{-1, 0, 1\}$. Note that 1-counter machines do not test the content of the counter. A configuration of the 1-counter machine is a pair in $Q \times Z$, that is, it consists of a state and the value of a counter. A transition relation $\rightarrow$ on configurations is defined as follows: $(q, z) \rightarrow (q', z')$ iff $(q, q', c) \in \delta_M$ and $z' = z + c$. The 1-counter 0-reachability problem is to decide whether there exists a state $q \in F_M$ such that $(q_0^M, 0) \rightarrow^* (q, 0)$. This is a special case of the empty-stack reachability problem for pushdown automata. While the latter is PTIME-complete, the following lemma shows that the former is in NLOGSPACE.

**Lemma 2.** *The 1-counter 0-reachability problem is in NLOGSPACE.*

*Proof.* Consider a 1-counter machine $M = (Q_M, \delta_M, q_0^M, F_M)$. We observe that for all $q, q' \in Q_M$, if there is a path $(q, 0) \rightarrow^* (q', 0)$, then there is such a path with stack depth bounded by $n^2$. This is a consequence of a summarization-based reachability algorithm (easily adapted from summarization-based reachability algorithm for pushdown automata), which computes summaries for pairs $(q, q')$. The iteration in which a pair $(q, q')$ gets added is the minimum absolute value of counter needed to reach from $(q, 0)$ to $(q', 0)$. The number of iterations is at most the number of summaries, that is, $n^2$. Note that this observation holds for all pushdown automata.

We can thus assume that the counter ranges over $(-n^2, n^2)$. State of a 1-counter machine is $(q, z)$, where $z$ is the value of a counter. Therefore we need to consider only $O(n^3)$ possible configurations. (This statement does not hold for general pushdown automata). Thus our reachability problem is a reachability problem in a graph with $O(n^3)$ states. The problem can be therefore solved in space $O(\log n^3)$.

**Theorem 1.** *The streaming data-string transducer equivalence problem is in* PSPACE.

*Proof.* Let us consider two streaming data-string transducers $S_1$ and $S_2$ from $\Sigma$ to $\Gamma$. They are not equivalent if there exists an input data string $w$ over $\Sigma$ such that one of the following three conditions hold: (i) $[\![S_1]\!](w)$ is defined, but $[\![S_2]\!](w)$ is not (or vice-versa), (ii) $[\![S_1]\!](w)$ and $[\![S_2]\!](w)$ are defined, but have different lengths, (iii) $[\![S_1]\!](w)$ and $[\![S_2]\!](w)$ are defined and have the same lengths, but there exists a position $p$ such that the data strings $[\![S_1]\!](w)$ and $[\![S_2]\!](w)$ differ at the position $p$.

We construct a 1-counter automaton and designate a state $q$ such that $q$ is 0-reachable in $M$ if and only if $S_1$ and $S_2$ are not equivalent. The automaton $M$ nondeterministically chooses which type of difference (of the three described above) it will find. We only describe here how $M$ can determine that there is an input string such that the $p$-th output symbol of $S_1$ is different from the $p$-th output symbol of $S_2$. The construction for the other two cases uses similar ideas and is simpler.

The automaton $M$ nondeterministically simulates $S_1$ and $S_2$ running in parallel. It keeps track of states of $S_1$ and $S_2$ precisely, but only keeps some information on the data and data string variables. Intuitively, $M$ guesses during the course of simulation of $S_1$ (resp. $S_2$) where the position $p$ in the output is, and uses its counter to check that the guess is the same for $S_1$ and $S_2$.

For each data string variable, $M$ guesses (at each step) where the contents of the variable will appear in the output with respect to the position $p$. More concretely, for each data string variable $x$ of both $S_1$ and $S_2$, $M$ guesses which of the following categories the variable is in: (i) left of $p$ (Class L), (ii) center, i.e. position $p$ is in this string (Class C), (iii) right of $p$ (Class R), (iv) $x$ does not contribute to the output (Class N).

Maintaining consistency of assignment of data string variables into these four classes is straightforward. First, consider the case when at a particular step, $S_1$ performs an assignment $y := (a, v_1)z(b, v_2)$ and $M$ guesses that the contents $y$ will appear to the left of the position $p$ in the output of $M_1$. To verify that this guess is consistent with previous guesses, $M$ checks that in the previous step, $z$ was in Class L. The assignment caused two output symbols $(a, v_1)$ and $(b, v_2)$ to appear to the left of the position $p$, therefore $M$ increases its counter by 2 (outputs of $S_2$ are taken into account by decreasing the counter rather than increasing). Second, if at a particular step, $S_1$ performs an assignment $x := (a, v_1)y(b, v_2)z$, and $M$ guesses that the symbol $(b, v_2)$ in this assignment will be at the position $p$, then: (i) $M$ verifies that at the previous step, $y$ was in Class L, and $z$ was in Class R, (ii) $M$ increases its counter by two in order to simulate the fact that $S_1$ outputs $(a, v_1)$ and $(b, v_2)$ (as before, when simulating $S_2$, $M$ decreases the counter), and (iii) $M$ assigns x to Class C. Note that initially, no variable is assigned to Class C, and at each step, at most one variable is in Class C, because of the copyless assignment restriction. The cases when $M$ guesses that a variable to which $S_1$ (resp. $S_2$) assigns is in Class R or Class N are similar.

In the remainder of the proof, we assume that the data domain $D$ is infinite. If it is finite, the automaton can directly store values from $D$ in its finite-state control, and the construction is simpler.

For data variables, $M$ keeps track of which variable is defined, and for the defined variables, it keeps track of the ordering and equality information. More precisely, let us consider the following set of variables $V = (V_1 \setminus \{curr^1\}) \cup (V_2 \setminus \{curr^2\}) \cup \{vp_1, vp_2, curr\}$, where $V_1$ and $V_2$ are the sets of data variables of $S_1$ and $S_2$, and $vp_1$ and $vp_2$ are used by $M$ to store information about the data value $S_1$ and $S_2$ output at position $p$. The automaton $M$ stores a pair $(V^d, \rho)$, where $V^d \subseteq V$ that contains all of the variables whose values are defined in computation of $S_1$ and $S_2$, and $\rho$ is an ec-order on $V^d$. The pair $(V^d, \rho)$ is updated as steps of $S_1$ and $S_2$ are simulated and their transitions are executed. Note that $M$ maintains only one variable $curr$ common to $M_1$ and $M_2$ because the two automata are running on the same input. The final part of the construction is the maintenance of $vp_1$, the variable used to store the output of $S_1$ at position $p$. When $M$ guesses that the output symbol of $S_1$ at position $p$ will be one in the right-hand side of the assignment (such as $x := (a, v_1)y(b, v_2)z$) it is simulating currently, it assigns $x$ to Class C as above, and if it guesses that at position $p$ is the symbol $(b, v_2)$, then : (i) the value $b$ from $\Sigma$ is stored in the finite state control of $M$, and (ii) $vp_1$ is added to $V^d$, the set of defined variables, and $vp_1$ is added to the equivalence class of $v_2$ in $\rho$. The construction for $vp_2$ is analogous.

To summarize, a state of $M$ consists of (1) a state of $S_1$, (2) a state of $S_2$, (3) a set $V^d \subseteq V$ representing the defined variables, (4) an ec-order $\rho$, (5) a partition of the data string variables of $S_1$ and $S_2$ to classes (as described above) and a symbol from $\Gamma$ at position $p$ for $S_1$ and $S_2$ (if $M$ guessed that the output to position $p$ was already performed). The set of states of $M$ is thus a product: $Q_1 \times Q_2 \times 2^V \times \rho \times Q_B$, with the components corresponding to items (1) to (5). The initial state of $M$ is the tuple containing initial states of $S_1$ and $S_2$, with the set $V^d$ empty — all the variables are undefined, and the component (5) has a special value $i$. From this state there are nondeterministic transitions which choose the initial assignments of data string variables to classes. The other transitions are as described above. The set of final states consists of states where either the variables $vp_1$ and $vp_2$ are defined, but $\rho$ does *not* imply $vp_1 = vp_2$ or the $\Gamma$ symbols stored in the finite-state control of $M$ for position $p$ in output strings of $S_1$ and $S_2$ differ.

We now prove that a final state of $M$ is 0-reachable iff there exists an input data string $w$ and a position $p$, such that $[\![S_1]\!](w)$ and $[\![S_2]\!](w)$ differ at position $p$. We will need the following notion that relates configurations of $M$ to configurations of $S_1$ and $S_2$. Let $c_1 = (q_1, \beta_1)$ be a configuration of $S_1$, let $c_2 = (q_2, \beta_2)$ be a configuration of $S_2$ and let $c_M = ((q_1^M, q_2^M, V^d, \rho, q_B), e)$ be a configuration of $M$ ($e$ is the value of the counter). The configuration $c_M$ is an abstraction of

configurations $(c_1, c_2)$ (denoted by $\alpha((c_1, c_2)) = c_M$) iff the following conditions hold:
- States: the states of $S_1$ and $S_2$ are the same in $c_1$ and $c_2$ as they are in $c_M$.
- Data variables: $\beta_1$ or $\beta_2$ are defined for each of the variables in $V^d$, and the values $\beta_1$ and $\beta_2$ assign to variables in $V^d$ are consistent with $\rho$
- Data string variables: Let $e_L^1$ be the number of symbols in the data string variables of $S_1$ that are assigned to Class L in $Q_B$. Class C in $Q_B$ contains by construction at most one data string variable of $S_1$. If Class C contains a data string variable $x_1$ of $S_1$, then we can designate a position $p_1$ in the data string in $x_1$. Let $e_C^1$ be the number of characters to the left of $p_1$ in $x_1$. The values $e_L^2$ and $e_C^2$ are defined analogously for $S_2$ and a position $p_2$ in a data string variable $x_2$. The following equality is required to hold: $e_L^1 + e_C^1 - (e_L^2 + e_C^2) = e$, where $e$ is the counter value in $c_M$. Furthermore, let $d_1$ be the data value at position $p_1$. We have that the equality and order relations that $\rho$ contains on $vp_1$ and the other data variables hold for $d_1$ and the values of these data variables given by $\beta_1$ and $\beta_2$. An analogous condition holds for the data value at position $p_2$.

*Claim 1* The automaton $M$ can reach the configuration $c_M = ((q_1, q_2, V^d, \rho, Q_B), e)$ in $k$ steps iff there exists an input string $w$ of length $k$, such that $S_1$ ($S_2$), after traversing this input, reaches a configuration $c_1$ ($c_2$), and $\alpha((c_1, c_2)) = c_M$.

The claim is proven by induction on $k$. The more difficult part of the proof of the claim is the left-to-right implication, where we are required to find an input string $w$ that satisfies the condition. We need to find a sequence of valuations $\beta$ that is the same as the sequence of pairs $(V^d, \rho)$ given by the sequence of configurations of $M$. It is here that Lemma 1 is used.

Using Claim 1, we now prove that a final state of $M$ is 0-reachable implies that there exists an input data string $w$ and a position $p$, such that $[\![S_1]\!](w)$ and $[\![S_2]\!](w)$ differ at position $p$. A final state of $M$ is a state where we do not have $vp_1 = vp_2$ or where the $\Gamma$ symbols stored for positions $p_1$ and $p_2$ differ. By Claim 1, this means that there is a position $p_1$ in the output of $S_1$ and a position $p_2$ in the output of $S_2$ where the data values or the $\Gamma$ symbols differ. If such a state is 0-reachable, (using Claim 1) we get that $e_L^1 + e_C^1 - (e_L^2 + e_C^2) = 0$, which implies $p_1 - p_2 = 0$, which implies that $p_1 = p_2$. The other implication can be also easily shown using Claim 1.

*Complexity* Checking whether a particular final state of $M$ is 0-reachable can be done in NLOGSPACE (Lemma 2). A nondeterministic algorithm first guesses which final state is reachable, and then checks its reachability in NLOGSPACE. The number of states of the 1-counter automaton $M$ we constructed is linear in the number of states of $S_1$ and $S_2$ and exponential in the number of data string and data variables of $S_1$ and $S_2$. Furthermore, given two states of $M$, one can decide (in polynomial time in the number of variables), whether there is a transition between the two states. We thus have that the streaming transducer equivalence problem is in PSPACE.

Theorem 1 implies that checking equivalence is in PSPACE for list-processing programs from Section 3 and list-processing functions defined in Section 4. The reason is that the number of data and output variables of the resulting transducer is linear in the size of the program (more precisely, in the number of data and reference variables of the program).

### 5.3 Checking pre/post conditions and assertions

Let $S$ be a streaming data string transducer $S$ from $\Sigma$ to $\Gamma$. Let $A_1$ be a streaming data string acceptor on $\Sigma$, and let $A_2$ be a streaming data string acceptor on $\Gamma$. The triple $\{A_1\}S\{A_2\}$ holds iff for all input data strings $w$ over $\Sigma$ we have that if $A_1$ accepts $w$ and $[\![S]\!](w) = w'$, then $A_2$ accepts $w'$. The *pre/post condition problem* for SDSTs is to determine, given $A_1$, $S$, and $A_2$, whether $\{A_1\}S\{A_2\}$ holds. Pre-post condition checking is useful in the context of verification, because we can, for example, ask whether a transducer that takes a sorted list (with respect to an ordering on $\Sigma$) as an input returns a sorted list (with respect to an ordering on $\Gamma$) as an output. The upper bound in the following theorem is obtained by reduction to the emptiness problem in nondeterministic finite automata (NFAs).

**Theorem 2.** *The pre/post condition problem for SDSTs is in* PSPACE.

The above definition of pre/post condition checking corresponds to partial correctness. We can also check total correctness: there is a PSPACE algorithm to check, given $S$, $A_1$, and $A_2$, is it the case that for all input strings $w$ accepted by $A_1$, $[\![S]\!](w)$ is defined and $A_2$ accepts $[\![S]\!](w)$.

The constructions discussed so far can also be used to solve a number of *assertion checking* problems for single-pass list-processing programs.

**Reachability** Given a single-pass list processing program $P$, a location $\ell$ of $P$, and a streaming data string acceptor $A$, is there a data string $w$ accepted by $A$ such that starting from the initial heap that stores $w$, there is an execution of $P$ leading to a configuration with location $\ell$? For this, we need to construct the SDST corresponding to $P$ as discussed in Section 3.4, and simulate it on an input together with $A$. The complexity is PSPACE. The same construction can be used if additional constraints are specified on boolean and tag variables of $P$ at the end of the execution.

**Pointer analysis** Given a single-pass list processing program $P$, two pointer variables $x$ and $y$, and a streaming data string acceptor $A$, is there a data string $w$ accepted by $A$ such that starting from the initial heap that stores $w$, there is an execution of $P$ leading to a configuration in which both $x$ and $y$ point to the same heap-node? Recall that the compilation of programs into SDSTs keeps track of such aliasing relationships, and has the necessary information to answer such a query. We can also check if a pointer variable $r$ is guaranteed to be non-null whenever it is dereferenced (using expressions such as $r.next$ and $r.data$).

**Heap-cycles detection** Given a single-pass list processing program $P$ and a streaming data string acceptor $A$, is there a data string $w$ accepted by $A$ such that starting from the initial (acyclic) heap that stores $w$, there is an execution of $P$ leading to a configuration in which the heap contains a cycle (formed by next-pointers of heap-nodes)? Again, the compilation of programs into SDSTs keeps track of the heap shape, and can be used to solve this problem in PSPACE.

### 5.4 Undecidable extensions

*Two-way data string transducers* A *two-way (deterministic) data string transducer* (2DST) is an extension of the streaming data string transducer model, where at each step, the transducer can decide whether to move left or to move right or to stay put. More precisely, a transition of a 2DST is defined by a a tuple $(q, \sigma, \varphi, q', \alpha, \zeta)$, where $q, \sigma, \varphi, q'$ and $\alpha$ are as for SDSTs, and $\zeta$ is in $\{\leftarrow, \downarrow, \rightarrow\}$. For 2DSTs, we assume that the input data string is enclosed by two special symbols $\vdash, \dashv$. The machine stops when it reaches a final state. If the machine never reaches a final state, or if it tries to move left while the tag is $\vdash$ or it tries to move right when the tag is $\dashv$, then the output undefined. Given a 2DST $S$ and a state $q$ of $S$, the *2DST reachability problem* is to determine whether there exists a data string $w$ such that $S$ enters the state $q$ while processing $w$.

**Theorem 3.** *The 2DST reachability problem is undecidable.*

The theorem is proven by reduction from the undecidable problem of emptiness for two-counter automata. The main step of the proof is to show that a 2DST can recognize whether the input data string encodes a computation of a two-counter machine. The proof uses the fact that the data domain is ordered.

*Programs with multiple traversing pointers* The class of imperative list processing programs considered in Section 3 restricts how next pointers of heap nodes can be traversed: there is one special pointer variable *curr*, and it is the only pointer variable that can traverse the next pointer. Now consider the class of programs, denoted by PMTP (short for programs with multiple traversal pointers), obtained by lifting this restriction, and allowing assignments $x := y.next$ for any two pointer variables $x$ and $y$. Given a program $P$ from the class PMTP and a location $\ell$, the *PMTP reachability problem* is to determine whether there exists a data string $w$ such that starting from the initial heap that stores $w$, there is an execution of $P$ leading to a configuration with location $\ell$.

**Theorem 4.** *The PMTP reachability problem is undecidable.*

The proof of the undecidability is again by a reduction from the reachability problem for two-counter automata. The basic observation is that if multiple pointers can traverse the heap simultaneously, the program can check whether two successive parts of the heap encode two successive configurations of the two-counter machine.

*Data string variable equality* While a number of analysis problems for SDSTs, and assertion checking problems for single-pass list-processing programs, are decidable, checking whether the transducer/program can reach a configuration where the contents of two string variables are the same, is undecidable. Given an SDST $S$, a state $q$ of $S$, and two data string variables $x$ and $y$ of $S$, the *data string variable equality problem* is to determine whether there exists a data string $w$ such that $S$ reaches a configuration where $x = y$ and the state is $q$. The following theorem is proven using a reduction from Post's correspondence problem.

**Theorem 5.** *The data string variable equality problem is undecidable.*

## 6  Related Work

We are not aware of any prior decidability results for checking semantic equivalence of list processing programs, even for the restricted case of bounded data domains.

The decidability of safety properties for programs with lists was investigated in [7]. The negative result in [7] holds for a very restricted class of programs: programs with only non-nested loops which do not modify the list data structure. Compared to the model of [7], we do not allow general traversal assignments of the form $x := y.next$, but allow only one pointer variable *curr* to traverse the next pointers of the heap nodes. We also assume that the initial heap is acyclic (but analysis algorithms can detect if cycles get introduced during program execution). In previous work [8], we have presented decidability results for a class of concurrent list accessing programs. The two models are different: the model in [8] allows concurrency and nondeterminism, but is not able to capture for example the list reversal transduction. The restrictions in [8] are rather intricate, and that is what triggered this study in search of a robust automata-theoretic model. Extending the streaming transducer model to capture concurrency is an interesting research direction. There is an emerging literature on automata and logics over data strings [15, 3] and algorithmic analysis of programs accessing data strings [2]. While existing literature studies acceptors and languages of data strings, we want to handle destructive methods that e.g. delete elements, and thus, a model of transducers is needed.

A number of automata-based techniques have been proposed for shape analysis [6, 5] (see also [9] for a survey). In particular, the *regular model checking* approach [6] employs transducers to reason about heap-manipulating programs in the following manner. The set of heaps feasible at a program point is represented by either a string automaton or a tree automaton, and the transformation on the heap due to a single statement is captured by a corresponding transducer model. The transformation of the entire program, then, corresponds to iterated composition of such transducers. Given a regular initial set of heaps, the set of heaps reachable after one transition will be regular. However, regular languages are not closed under unbounded union, so the set of all reachable heaps need not be regular. Consider a program that given an input list $w$ outputs the list

$ww$ (note that data values do not play an important role in this transduction). For such a program, the iterative fixpoint procedure to compute the set of all reachable configurations does not terminate (in fact, the set of reachable configurations is not regular, and cannot be represented by a finite-state automaton). However, a streaming transducer that computes such a transduction can be easily defined. It is important to note that our decision procedures do not attempt to compute the set of reachable configurations (or heap contents). The literature on regular model checking provides several techniques for over-approximations of the set of reachable heaps to ensure termination, such as widening [20] and specialized abstractions using counters [4].

Analyzing programs that manipulate dynamic linked data structures is a widely studied problem commonly described as *shape analysis* [19]. Proving assertions of such programs is undecidable [12, 18], and the bulk of the literature consists of abstraction-based techniques for verification (see e.g. [17, 13, 10]). The core problem these techniques focus on is computation of invariants that often need to quantify over the nodes in the heap. Let us consider function `Delete` from Section 3. If the function was called with the parameter $d$, a natural postcondition is that all nodes reachable from `result` have values different from $d$. A quantified invariant needed to prove the postcondition could be automatically computed using for example the approach described in [17]. However, we emphasize that in contrast to the sound-but-not-complete abstraction-based methods for checking safety properties, our approach is sound and complete for a well-defined class of programs and, in addition to checking of assertions and pre/post conditions, we presented an algorithm for checking equivalence of programs.

## 7 Conclusions

We have introduced a streaming transducer model, and showed that it can serve as a foundational model of single-pass list processing programs. Our results lead to algorithms for checking functional equivalence of two programs, written possibly in different programming styles, for commonly used routines for processing lists of data items. We are not aware of any prior decidability results for checking semantic equivalence of list processing programs, even for the restricted case of bounded data domains.

We also believe that the streaming transducer model introduced in this paper is of independent theoretical interest. We have started the investigation of expressiveness and related theoretical properties of the transducer model when the data domain is bounded. Classical string-to-string finite-state transducers need to be "two-way" to implement an operation such as reverse. In a subsequent paper, we showed that the streaming string transducer model is expressively equivalent to two-way transducers [1], and thus, to MSO-definable string transductions [11]. Learning streaming string transducers from input/output examples, and defining a similar streaming transducer model for tree-structured data are potential fruitful directions for future research.